# Cartographier des trajectoires maritimes incertaines du XVIII[ème] siècle

**Christine Plumejeaud-Perreau[1], Bernard Pradines[1]**

*1. UMR 7301 MIGRINTER, CNRS & Université de Poitiers*
  *5 rue des Theodore Lefebvre, 86073 Poitiers, France*
  *[prenom.nom@univ-poitiers.fr](prenom.nom@univ-poitiers.fr)*

*RÉSUMÉ. Cet article présente comment des trajectoires de navires ont été reconstruites à partir de sources historiques concernant le commerce maritime du XVIIIème siècle. Il s'intéresse dans un premier temps à la qualification de l'incertitude inhérente à ces sources, en expliquant la méthode de reconstitution des itinéraires. Il détaille ensuite comment la géométrie des tronçons de trajectoire reliant les escales par la mer ont automatiquement été calculés en vue de leur cartographie. L'algorithme programmé en PL/SQL est transposable à d'autres contextes de recherche et disponible sous licence libre. Il expose enfin un outil de recherche et géovisualisation sur le Web de ces trajectoires avec leur niveau d'incertitude associé et son utilité pour les historiens, mais aussi pour l'analyse et la vérification des trajectoires ainsi reconstruites.*

*ABSTRACT. This article presents how ship'trajectories have been built from historical sources dealing with maritime trade in the 18th century. It first summarizes the method for building the routes, and qualifying the uncertainty level linked to each of its segments. Then, it details how the geometries of these segments connecting two successive stopovers were automatically calculated in order to draw a map with maritim paths only. The algorithm, programmed with PL/SQL language, is available under an open-source licence ([https://gitlab.huma-num.fr/portic/porticapi](https://gitlab.huma-num.fr/portic/porticapi)). Finally, an online tool for querying and mapping these routes with their associated level of uncertainty is exposed (http://shiproutes.portic.fr). We show its usefulness for historians, in particular for the control of the validity of built ship'trajectories.*

*MOTS-CLES : Cartographie de trajectoire, calcul de chemin, incertitude, géovisualisation*

**1. Introduction**

Cette recherche s'intéresse à l'incertitude des données historiques portant sur le commerce maritime au XVIIIème siècle afin d'en proposer une visualisation expressive et en particulier cartographique. En effet, nous étudions des sources saisies dans une base de données spatio-temporelle qui nous permettent de reconstituer au moins une année de circuits de navigation sur les côtes françaises, du ponant au levant. L'apport en termes d'analyse est considérable - par exemple pour déterminer les aires de navigation en fonction du tonnage ou encore la durée des rotations - à la condition de lever ou d'expliciter les incertitudes inhérentes à cette reconstitution. Or, dans sa volonté de proposer une visualisation expressive et en particulier cartographique des données sur la navigation, l'équipe interdisciplinaire a fait le choix de donner à voir cette incertitude, ce qui requiert un travail complexe de curation des données et de qualification des degrés d'incertitude des principales variables.

La base de données contient, entre autres, la transcription numérisée de plus de 30 000 congés de navires dans les ports français en 1787. Un congé est une déclaration fiscale d'un capitaine de navire à l'Amirauté d'un port qui l'autorise à partir. En outre, cette base de données contient deux autres sources, à savoir le "Registre sanitaire de Marseille" et les "Cahiers du petit cabotage". Certaines descriptions de navires correspondent entre ces sources et peuvent confirmer une déclaration de l'autre source. Cette vérification est rendue possible par l'identification de navires semblables à travers ces différentes sources.

Cet article élicite d'abord la nature de l'incertitude que nous souhaitons représenter, à partir d'exemples concrets. L'information qui est traitée est de nature très imparfaite (Devillers et Jeansoulin 2005) du fait d'informations inégalement disponibles, de l'imprécision de la formulation des lieux ou des dates, mais également d'informations fausses ou contradictoires. Cette communication résume les règles de l'algorithme qui permet de contrôler la cohérence des informations, (Plumejeaud-Perreau et Marzagalli, 2022), afin de se focaliser sur la cartographie des itinéraires maritimes. Nous proposons un outil de géovisualisation permettant de rechercher et cartographier ces itinéraires, lesquels sont calculés automatiquement par des chemins maritimes suivant une méthode inédite.

Notre article est organisé suivant trois parties. La première résume la méthode de de qualification automatique de l'incertitude inhérente à ces hypothétiques itinéraires. La deuxième partie détaille comment les géométries de tronçons reliant deux escales successives par la mer sont automatiquement calculées en vue de leur cartographie. La troisième partie présente un outil de recherche et géovisualisation de ces trajectoires avec leur niveau d'incertitude associé et son utilité pour les historiens.

**2. Reconstitution de trajectoires à partir de sources historiques**

Cette section décrit comment les trajectoires de navires peuvent être reconstituées à partir d'une base collectant des déclarations d'itinéraires, de sources hétérogènes, dont deux majeures qui sont l'une, les congés fiscaux des Amirautés de France pour 1787 et 1789 qui constituent une partie importante (67 %) de la base de données, et l'autre les recensements sanitaires à l'arrivée à Marseille des navires pour les années 1749, 1759, 1769, 1779, 1787, 1789, 1799.

Dans les congés, les capitaines déclarent au lieu de départ une intention de destination, alors que dans le registre sanitaire de Marseille, ce sont les escales précédant l'arrivée à Marseille qui sont connues avec certitude. La base de données qui enregistre les sources est structurée autour de la notion centrale de point d'escale, qui est un événement se déroulant en un lieu à un instant. En dépit de l'imprécision de la mention du lieu d'escale parfois, celui-ci est rapporté à un des lieux référencés dans notre index géographique, le plus souvent un port mais pas toujours (un naufrage au large des côtes, une destination vague comme « Côte de Bretagne »), en notant son identifiant en plus de la variante toponymique lue dans la source. Cet index a aussi l'avantage de préciser l'entité géographique administrative d'appartenance de 1787, Etat, province, Amirauté ou autre (Plumejeaud et al., 2021). Concernant la date de l'escale, elle est renseignée soit comme une date de sortie ou une date d'entrée, en fonction du mouvement du navire, et elle est précise au jour près sur le lieu d'observation. Les dates sont dans un format spécial : par exemple, 1787=05=10 si exactement à la date du 10 mai 1787, 1787>05>10 si après cette date, ou 1787<05<10 si avant cette date. Ce format spécifie ainsi l'incertitude sur la date d'arrivée à une destination dans les congés, ou de sortie d'une escale précédent l'arrivée à Marseille. Il autorise un ordonnancement lexicographique équivalent à l'ordre temporel, et peut être converti en format temporel normé ISO 8601 en remplaçant les caractères spéciaux (=, < ou >) par des tirets.

Tout document source, qui correspond à une ligne ou un feuillet de déclaration par une même personne dans un port un certain jour, est retranscrit dans une table centrale, nommée « *pointcall* ». Cette table décrit les escales point par point réalisées par un navire, piloté par un capitaine, suivant un document source auquel la base attribue un identifiant (*data_block_local_id*). Par exemple, pour un document référencé « 00188143» dans les sources décrivant le trajet d'un navire du Havre à Marseille en passant par le Détroit de Gibraltar, la relation *pointcall* contient 3 entrées, décrivant l'action (*pointcall_action*) réalisée par le navire en ces trois ports qui peut être grosso modo soit d'arriver, soit de partir, soit de partir pêcher dans le voisinage côtier, soit de faire un transit. La saisie des informations communes (nom du navire et du capitaine, tonnage, type de navire, pavillon, port d'attache parfois, origine du capitaine, taille de l'équipage, etc.) se fait au point d'observation, celui où est enregistré la déclaration dans le document source, qui est marqué par un attribut *pointcall_function* égal à 'O'. Ailleurs, la description est recopiée automatiquement. Un attribut *pointcall_status* explicite si d'après le document, l'évènement a vraiment eu lieu dans le passé (il a un statut 'PC') ou bien si c'est une intention annoncée (statut 'FC') ou bien si c'est une intention qui n'a pas pu être réalisée (statut 'PU').

Cette information est essentielle puisqu'elle va permettre de rendre compte d'incertitudes historiques. Peut-on en effet confirmer par le croisement de sources qu'une intention (statut 'FC') s'est vu réalisée ultérieurement ? L'ordre de visite de l'escale pour le navire dans ce document est renseigné manuellement dans *pointcall_rank*. Il va permettre d'ordonner les escales dans l'ordre chronologique, plus facilement que les dates indiquées dans l'un (et/ou l'autre) des attributs pointcall_out_date et pointcall_in_date. Ceci résume ici quelques informations essentielles à la compréhension de la transcription des sources dans la base, décrite par ailleurs (Dedieu et al., 2011).

L'information peut faire l'objet plusieurs interprétations qui vont être ajoutées sur la ligne saisie immédiatement ou *a posteriori* : reconnaître et identifier un même capitaine et lui attribuer un identifiant (*captain_id*), ou un navire déjà repéré dans d'autres documents source et l'identifier par un numéro unique (*ship_id*) : 13 820 navires différents ont été identifiés à ce jour. Notre étude écarte 28 % des 65 030 documents n'ont pas encore fait l'objet de cette identification. Pour la suite, nous allons supposer que cette identification est fiable, et nous concentrer sur les différents types d'incertitude relatifs aux déclarations du mouvement des navires dans le temps (désignés comme chemins ou itinéraires).

### *2.1. Vérification automatique de l'enchaînement des escales*

Toutes les sources ne présentent pas le même statut du point de vue de la réalité des faits énoncés. Ainsi, les sources marseillaises concernent des arrivées, donc des événements passés au moment du récit aux officiers de santé (*pointcall_status* = PC) : le registre mentionne le lieu et la date exacte d'où le navire a appareillé, ainsi que tous les endroits où il a fait escale avant Marseille, certifiés par un visa daté et tamponné. Tous ces lieux sont qualifiés de « observés », *i.e.* certains, dans la variable *pointcall_uncertainty*. Dans quelques cas, les registres de Marseille mentionnent la prochaine destination prévue. Comme pour tous les événements futurs (*pointcall_status* = FC), il est possible que cette intention n'ait pas été réalisée et le pointcall est qualifié de « déclaré ». Dans le cas des congés, le lieu de prise du congé est « observé », donc certain, tandis que les destinations annoncées sont qualifiées *a priori* de « déclarées ». Il arrive parfois qu'un capitaine indique sa précédente escale (*pointcall_status* = PC) en prenant son congé, mais elle n'est pas certaine car c'est une déclaration, qualifiée comme telle : « déclarée ».

La reconstitution d'itinéraire peut se faire en ordonnant chronologiquement les escales d'un même navire. C'est le cas par exemple du navire « la Fidèle Mariane » suivi à travers 6 documents, dont les déclarations sont toutes parfaitement cohérentes : *pointcall_uncertainty* prend la valeur de « confirmé » pour les intentions d'escales en vert dans le tableau 1 qui sont confirmées par l'observation consécutive. Quand les dires sont concordants, il y a un doublon au point d'arrivée et de départ, comme pour la Fidèle Mariane. Pour reconstituer l'itinéraire, il faudrait simplement éliminer un des doublons, en gardant si possible l'escale précisément datée (donc avec *pointcall_function* à 'O'). C'est un des rôles dévolus à la variable « *net_route_marker* » introduite par les historiens, lorsqu'elle prend la valeur 'A' (pour conserver l'escale), ou 'Z' (pour éliminer le doublon).

TABLE 1. *Ordonnancement des escales du Fidèle Marianne, et confirmation de ses déclarations d'intention.*

| data_block_local_id | ship_id | ship_name | pointcall | out_date | rank | net_route_marker | function | status |
|---|---|---|---|---|---|---|---|---|
| 00062213 | 0002931N | Fidèle Mariane | Les Sables-d'Olonne | 1787=01=05 | 1 | A | O | PC |
| 00062213 | 0002931N | Fidèle Mariane | Bayonne | | 2 | Z | T | FC |
| 00140197 | 0002931N | Fidèle Mariane | Bayonne | 1787=03=16 | 1 | A | O | PC |
| 00140197 | 0002931N | Fidèle Mariane | Dunkerque | | 2 | Z | T | FC |
| 00110541 | 0002931N | Fidèle Mariane | Dunkerque | 1787=05=10 | 1 | A | O | PC |
| 00110541 | 0002931N | Fidèle Mariane | Les Sables-d'Olonne | | 2 | Z | T | FC |
| 00102845 | 0002931N | Fidèle Mariane | Les Sables-d'Olonne | 1787=06=18 | 1 | A | O | PC |
| 00102845 | 0002931N | Fidèle Mariane | Bayonne | | 2 | Z | T | FC |
| 00140566 | 0002931N | Fidèle Mariane | Bayonne | 1787=09=04 | 1 | A | O | PC |
| 00140566 | 0002931N | Fidèle Mariane | Saint-Malo | | 2 | Z | T | FC |
| 00142100 | 0002931N | Fidèle Mariane | Saint-Malo | 1787=10=08 | 1 | A | O | PC |
| 00142100 | 0002931N | Fidèle Mariane | Saint-Brieuc | 1787<10<10! | 2 | A | | FC |
| 00142100 | 0002931N | Fidèle Mariane | Côtes de Bretagne | | 3 | A | T | FC |

TABLE 2. *Ordonnancement des escales du Suzanne, avec croisement des sources de Marseille et des congés de l'Amirauté.*

| data_block_local_id | ship_id | ship_name | pointcall | out_date | in_date | rank | function | Net_Route_marker | status |
|---|---|---|---|---|---|---|---|---|---|
| 00294615 | 0012925N | Suzanne | La Rochelle | 1787=06=01 | None | 1 | O | A | PC |
| 00294615 | 0012925N | Suzanne | Seudre | None | 1787>06>01! | 2 | T | A | FC |
| 00149798 | 0012925N | Suzanne | La Tremblade | None | 1787=06=05 | 1 | O | A | PC |
| 00151273 | 0012925N | Suzanne | Marennes | 1787=06=19 | None | 1 | O | A | PC |
| 00151273 | 0012925N | Suzanne | Le Havre | None | 1787>06>19! | 2 | T | Z | FC |
| 00162284 | 0012925N | Suzanne | Le Havre | 1787=08=04 | None | 1 | O | A | PC |
| 00162284 | 0012925N | Suzanne | Marseille | None | 1787>08>04! | 2 | T | Z | FC |
| 00188143 | 0012925N | Suzanne | Le Havre | 1787=08=06 | None | 1 | A | Z | PC |
| 00188143 | 0012925N | Suzanne | Détroit de Gibraltar | None | 1787=08=24 | 2 | None | historien Z / algorithme A | PC |
| 00188143 | 0012925N | Suzanne | Marseille | None | 1787=09=14 | 3 | O | A | PC |

Mais parfois, comme à Marseille, on retrouve des navires qui ont pris congé sur l'atlantique par exemple en annonçant leur intention future d'aller à Marseille. A Marseille, lorsqu'on les retrouve, les étapes détaillées depuis ce port atlantique sont détaillées comme pour le navire Suzanne (tableau 2) : parti du Havre le 6 aout 1787, alors qu'il a pris son congé le 4 aout 1787, le navire a passé le détroit de Gibraltar le 24 aout 1787 avant d'arriver à Marseille le 14 septembre 1787. Ainsi des intentions d'escales s'intercalent entre les vraies escales dans l'ordre chronologique et peuvent tromper la simple logique de l'ordonnancement chronologique : le Suzanne n'a pas navigué du Havre à Marseille puis au Havre en deux jours. L'insertion d'un Z devant ces escales « intentionnelles » (déclarées comme FC dans *pointcall_status*) permet d'éviter cela.

Le Z du *net_route_marker* prend également souvent une autre signification, celle de discréditer l'intention annoncée qui n'a pas pu se réaliser, et parfois suivant le point de vue de l'historien. Il y a plusieurs cas qui correspondent à cette situation :

(i) L'intervalle de temps entre les escales ne permet pas la réalisation matérielle des tronçons. C'est le cas par exemple du Turgot, 0021517N, qui quitte Nantes le 4 juillet 1787 pour aller à la Guadeloupe, mais qu'on retrouve à Bordeaux le 12 octobre 1787 déclarant partir pour la Guadeloupe. Sachant qu'à l'époque, un trajet vers les îles antillaises comprend une pause de trois mois aux îles, cela semble impossible : en réalité, depuis Nantes, le navire est passé par Bordeaux avant de filer vers la Guadeloupe. L'analyse intègre ici l'expertise d'historien, car les statistiques que nous avons produites sur la durée moyenne d'un trajet sont extrêmement variables, car dépendantes de la météo et les conditions de navigation de l'époque.

(ii) On démontre que c'est une fausse déclaration (*pointcall_uncertainty* vaut « invalidé ») parce qu'on ne voit pas le navire ressortir de la destination mentionnée, alors que registre des sorties de cette destination annoncée est disponible (Marzagalli et Pfister, 2014), et que le même navire est observé ultérieurement ailleurs.

Quant aux pointcalls succédant au dernier point d'observation d'une suite de documents, ils restent « invérifiables » comme les destinations vers Saint Brieuc et « Côtes de Bretagne » du Fidèle Marianne. Ce qui est également le cas de navires qui n'apparaissent que dans un seul document source, ou de destinations vers l'étranger ou des zones sans registres (comme La Guadeloupe visitée par le Turgot).

Sur la base des règles énoncées, un algorithme (Plumejeaud et Marzagalli, 2022) met à jour la valeur de *net_route_marker* et de *pointcall_uncertaint.* avec ses six niveaux, à savoir : observé - confirmé - déclaré - controversé - invérifiable - invalidé. Comme la règle (i) concernant l'intervalle de temps est impossible à évaluer, l'algorithme indique généralement 'A' pour garder le point. Si la valeur de *net_route_marker* diffère de la proposition de l'historien, alors le pointcall est qualifié de « controversé ». Cela signifie qu'il faut se pencher attentivement sur le cas de ce pointcall pour vérifier qui se trompe, de l'algorithme ou l'historien.

*2.2. Qualification de l'incertitude des tronçons composant un itinéraire*

Afin de construire l'itinéraire, l'algorithme joint les escales de départ aux arrivées ordonnées chronologiquement par navire identifié, en éliminant les points Z sauf ceux controversés. Il reconstruit à la fois les tronçons déclarés dans les sources, reconnus comme tels par la variable « direct » valuée à vrai, ainsi que les retours implicites déduits (« direct » vaut faux) qui ont dû se faire, comme le trajet de la Tremblade à Marennes de la Suzanne entre le 05 juin 1787 et le 19 juin 1787. Ces tronçons reconstruits représentent 21 % des 70 990 tronçons qualifiés par notre approche. Chaque tronçon est qualifié par une valeur ordonnée d'incertitude (*travel_uncertainty*), déduite à partir de l'incertitude de chaque extrémité (table 3), sachant que les extrémités d'incertitude plus forte sont prioritaires pour décider du niveau d'incertitude d'un tronçon.

TABLE 3. Vue synthétique sur les niveaux de certitude qualifiée de chaque tronçon de navigation.

| Travel_ uncertainty | Signification | pointcall_uncertainty sur le départ et/ou l'arrivée | Nombre | Couleur |
|---|---|---|---|---|
| 0 | Confirmé | confirmé ou observé, aux deux extrêmités | 35 355 | Vert |
| -1 | Déclaré | déclaré, invérifiable à l'une des extrémités | 28 768 | Gris |
| -2 | Invalidé | Invalidé à l'une des extrémités | 5 544 | Rouge |
| -3 | Controversé | controversé à l'une des extrémités | 1 323 | Orange |

Nous pensons avoir qualifié ainsi de façon réfutable et fiable l'incertitude liée à chaque escale dans une grande majorité de cas. Il reste cependant possible que la visualisation des trajectoires des navires dévoile d'autres corrections nécessaires de l'algorithme, ou amènent les historiens à réviser leur jugement. Pour ce type d'exploration analytique, au cas par cas, il nous a semblé pertinent de l'associer à une carte (Ducruet, 2013). Mais nous souhaitions cette cartographie plus agréable à lire que celle de tronçons coupant en droite ligne des terres, comme le trait noir de la figure 1 reliant Saint-Malo à Saint-Brieuc.

**3. Calcul de géométries reliant des escales par voie de mer**

Pour calculer des géométries de tronçons qui suivent des voies maritimes, nous souhaitions une approche simple et reproductible, utilisable en dehors du contexte de cette recherche historique et qui n'exploite pas de sources historiques comme peut le faire le projet CLIWOC[1]. Celui-ci exploite en effet les journaux de bord des capitaines pour retracer les grandes routes maritimes empruntés par les navires de la période couvrant 1750 à 1850. Notre approche est celle d'un calcul automatique d'évitement d'obstacle (*pathfinding*) sous contraintes, et s'inspire des travaux de (Poncet-Montanges, 2013). Si son code déposé sur une drop-box reste inaccessible aujourd'hui, son rapport dessine le fonctionnement dans les grandes lignes de son algorithme qui, comme nous, n'est pas un algorithme de calcul de plus court chemin sur graphe En effet, les techniques de recherche opérationnelle sur graphe (Chauveau, 2017) requièrent généralement une discrétisation prédéfinie de l'espace de recherche, alors que l'océan est un espace continu et qu'il n'y a pas de manière directe de le mailler (Dupuy, 2021). Souvent ces approches mobilisent un grand nombre de paramètres (météo, type de navire, danger de capture en mer, etc.) que nous souhaitions écarter pour plus de simplicité.

---

[1] http://www.ucm.es/info/cliwoc/

Le seul paramètre qui nous est nécessaire est l'expression de la contrainte, celle d'un trait de côte que le tronçon ne doit pas intersecter. Nous avons utilisé celui délivré par l'agence nationale américaine des océans et de l'atmosphère[2], sous la forme d'un shapefile nommé « GSHHS_i_L1.shp » qui est extrait de la base « *Global Self-consistent, Hierarchical, High-resolution Geography Database* » (GSHHG) dans sa version 2.3.7 (Wessel et Smith, 1996). Sa résolution est dite intermédiaire « i » selon l'agence, ce qui signifie généralisée suivant l'algorithme de Douglas-Peucker à un niveau de 80% par deux fois à partir du trait de plus haute résolution disponible, et projeté en WGS84. Ce choix nous permet de voir les contours des petites îles souvent citées dans nos sources comme l'île de Ré, l'île de Bouin, tout en optimisant les ressources nécessaires au calcul.

En effet, notre calcul dessine une ligne entre le début et la fin d'un tronçon, mais la partie du segment intersectant la côte est extraite pour construire un tampon le long de la côte, pour lequel un ensemble de points intermédiaires sont placés à espaces réguliers. Le croquis (fig. 1) expose son fonctionnement dans le cas relativement simple du tracé de la route maritime de Saint-Malo à Saint-Brieuc.

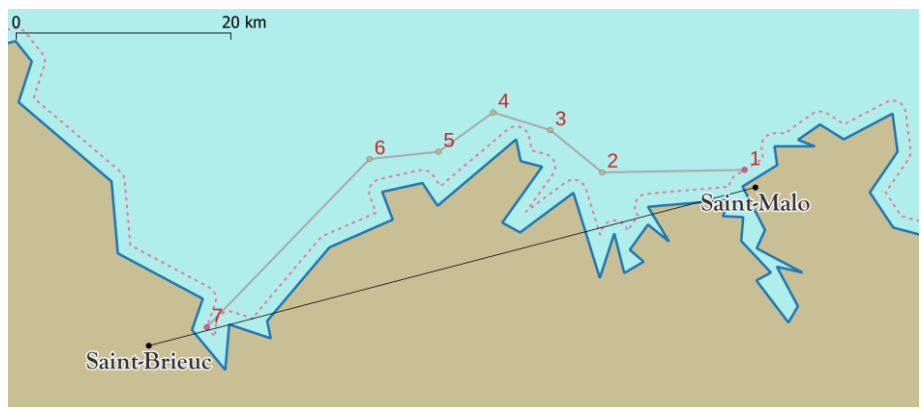

FIGURE *1. Tracé de Saint-Malo à Saint-Brieuc avec ses 7 points intermédiaires.*

---

2 Site Web d'accès aux données et métadonnées de la NOAA :
https://www.ngdc.noaa.gov/mgg/shorelines/shorelines.html

La largeur du tampon (*offset*) ainsi que l'espacement entre les points (*distance_ratio*) sont des paramètres du programme qui s'ajustent automatiquement en fonction des distances à parcourir. Par exemple, sur Saint-Malo à Saint-Brieuc, le tampon est de 5 km et l'intervalle entre les points est de 10 km, mais entre Saint-Brieuc et les Côtes de Bretagne, ils augmentent respectivement à 20 km et 20 km. Ceci correspond à un souci d'économie de ressources, car en passant plus au large, il faut calculer moins d'intersections de côtes et moins de points. De même, les îles de moins de 1 km$^2$ sont traversées, comme l'île de Béniguet dans la figure 2. Cette approche correspond aux pratiques de navigation à voile depuis l'antiquité jusqu'à la fin de l'époque moderne : naviguer à vue le plus longtemps possible le long des côtes, avant de s'élancer en droiture sur de vastes étendues océanes seulement si nécessaire, en ligne droite suivant un cap pour rejoindre un amer proche de la destination et à nouveau caboter (Bochaca et Moal, 2019). L'invariant du programme est qu'il n'existe aucune intersection entre la terre et un tronçon reliant deux points intermédiaires ou finaux. Ceci explique que sur la figure 1, le tracé n'est pas réalisé des ports exactement (points noirs), mais depuis leur projection (points roses n° 1 et n°7) sur une ligne située à 1 mile nautique en pointillés rose (*i.e.* 1,852 km, soit la portée maximale d'un boulet de canon de l'époque).

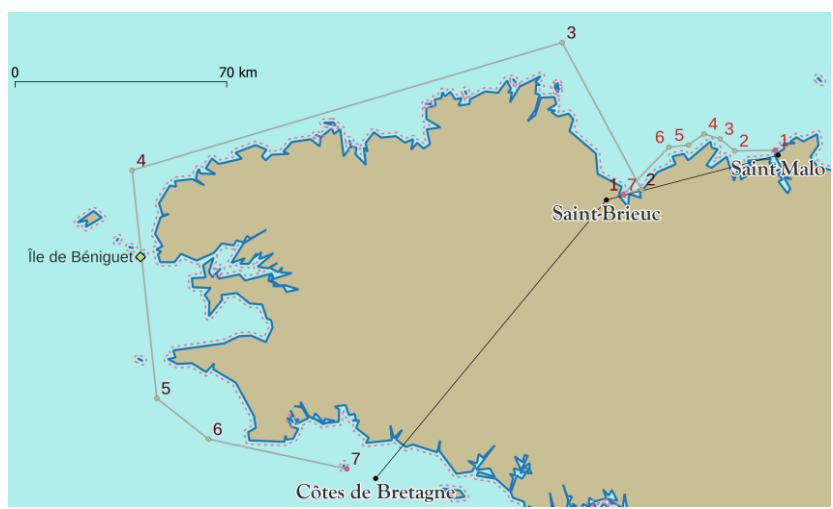

FIGURE 2. *Tracé à Saint-Brieuc à Côtes de Bretagne.*

L'algorithme fonctionne récursivement afin de dépasser la difficulté liée à la forme fractale des côtes, mais également pour contourner les îlots qui se trouvent très souvent à l'entrée de zones portuaires. La figure 3 illustre comment le programme repousse automatiquement un point en mer lorsqu'un des points intermédiaires tombe malencontreusement sur une île. Le mauvais point P3 est remplacé par son symétrique P'3 par rapport à la côte. Ce tirage aléatoire réussit le plus souvent, mais sinon, l'algorithme reboucle pour déplacer P'3 jusqu'à ce qu'il soit placé en mer.

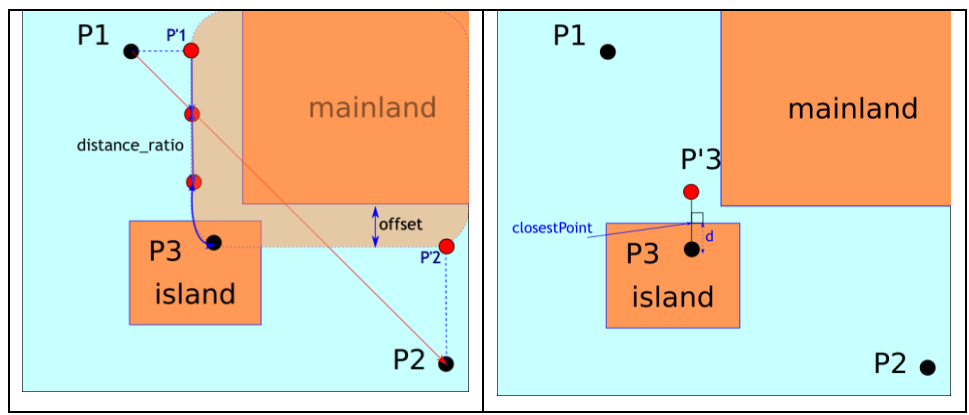

FIGURE *3. Déplacement du point P3 intermédiaire à P'3 pour éviter la terre.*

L'itération est ainsi récursive, et permet de calculer un ensemble de routes maritimes qui n'intersectent jamais les parties côtières. Le chemin tracé fait ensuite l'objet d'une simplification, suivant un algorithme de complexité n log(n) afin de supprimer les boucles formées par le tirage de points hors mer. Le calcul est implémenté sous la forme d'une procédure PL/SQL[3] qui renvoie une structure de données comprenant la liste ordonnée des points, mais également les valeurs ajustées des paramètres (tampon et espacement des points), ainsi que la durée du calcul en millisecondes. Sur les 7000 tronçons calculés, la durée moyenne varie entre 80 et 500 secondes, ceci en fonction de la distance à parcourir en ligne droite (trait noir sur les figures 1 et 2), qui conditionne les valeurs de tampon et d'espacement entre les points. Ce résultat nous permet de proposer d'explorer les parcours des navires et des capitaines via une cartographie interactive sur le Web.

**4. Géovisualisation interactive de trajectoires de navires ou de capitaines**

Cet outil de visualisation, disponible sur le Web[4], permet de rechercher des itinéraires par le nom ou l'identifiant soit d'un capitaine soit d'un navire. La sélection peut s'affiner en fonction de différents descripteurs, tels que le pavillon, le port d'attache, le tonnage du navire ou les dates de navigation. Lorsqu'un itinéraire a été choisi (ou deux au maximum), l'affichage (fig. 4) comprend deux vues qui se complètent pour faire comprendre le scénario reconstruit pour l'itinéraire avec ses différentes alternatives.

---

[3] https://gitlab.huma-num.fr/portic/porticapi/-
  /blob/master/sql/portic_routepaths_allfunctions.sql

[4] http://shiproutes.portic.fr

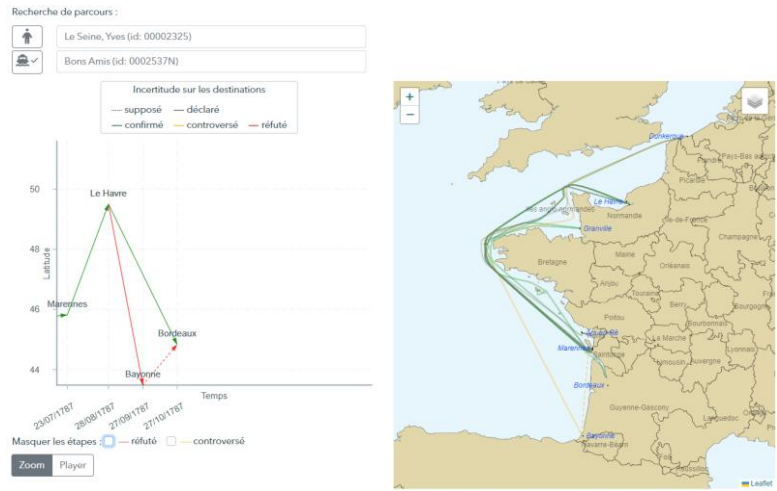

FIGURE *4. Affichage du chronogramme et carte de l'itinéraire du navire Bons Amis, (identifiant 0002537N), piloté par Le Seine, Yves dans* http://shiproutes.portic.fr

A gauche de l'écran, un chronogramme déroule l'enchaînement des étapes dans le temps. L'axe vertical dispose les escales en fonction de leur latitude, l'axe horizontal les dispose en fonction de la date de leur réalisation. Il s'inspire de travaux de recherche récents sur les récits de migrations (Dubucs et al., 2010), (Bahoken, 2021) et les manières traditionnelles d'afficher les itinéraires des navires (Buti, 2003). En effet, l'existence de nombreux allers-retours entre les mêmes ports, fréquents à l'époque, nuit à la lisibilité des cartes, et l'interactivité permet de résoudre ce souci (Mac Eachren, 2005). Le chronogramme est interactif : en zoomant dans cette partie, les dates des escales très rapprochées peuvent être distinguées, comme pour les Bons Amis piloté par Le Seine, Yves (fig. 4), qui est parti de Marennes le 23 juillet 1787 pour le Havre. Ce chronogramme est associé à un outil de déroulement interactif, étape par étape qui déclenche la visualisation des détails de l'itinéraire dans la partie droite de l'écran, dédiée à la carte interactive.

La sémiologie pour les tronçons est commune à ces deux vues : elle vise à rendre compte de l'incertitude calculée des tronçons d'itinéraire (cf. tableau 4). La teinte d'une palette divergente est utilisée pour montrer le niveau de certitude (du vert pour « confirmé » au rouge pour « invalidé ») tandis que les lignes pointillées sont utilisées pour montrer les étapes qui ont été inférées par l'algorithme. Nos choix ont également été guidés par la littérature scientifique sur les signes d'incertitude (Arnaud et Davoine, 2009). Ainsi, la route calculée peut être cartographiée soit dans sa globalité comme sur la figure 4, soit dans le détail comme dans la figure 5. Par exemple, le navire les Bons Amis n'a pas réalisé le trajet du Havre à Bayonne comme il le déclarait, qui aurait impliqué un retour à Bordeaux (en rouge pointillés), mais seulement un trajet du Havre à Bordeaux, confirmé en vert. Sur chaque tronçon, le zoom de la carte s'adapte automatiquement, et une bulle au survol de la souris affiche les détails de l'armement du navire sur ce tronçon.

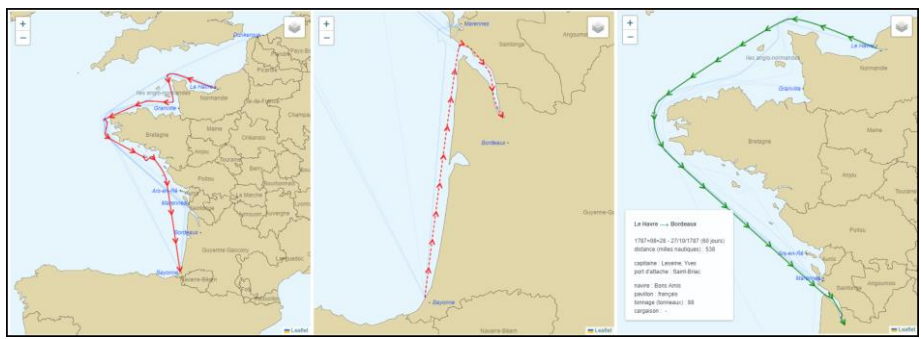

FIGURE 5. *Déroulement de l'itinéraire dans la partie cartographique du navire Bons Amis, (identifiant 0002537N), dans http://shiproutes.portic.fr.*

La finalité de l'application est double. D'une part, elle permet pour des passionnés d'histoire maritime de retrouver des personnages ou des navires cités dans d'autres sources, et nous avons un exemple avec ce marin, Pierre-Etienne Millié dont la biographie est commentée dans une revue locale d'histoire maritime (Millié, 2021), qui cite le navire « Comte de Forcalquier » mais sous une autre orthographe – « Comtede Forcalyner » - identifiant 0009151N. D'autre part, elle permet de vérifier visuellement les cas controversés comme l'arrêt à Gibraltar de la Suzanne (tab. 2), mais aussi de vérifier si l'identification de navires est cohérente, comme dans le cas suivant (fig. 6) présentant les trajets de deux navires distincts. Développée comme un client Web riche avec *React*, l'outil exploite l'API en ligne[5].

---

[5] http://data.portic.fr

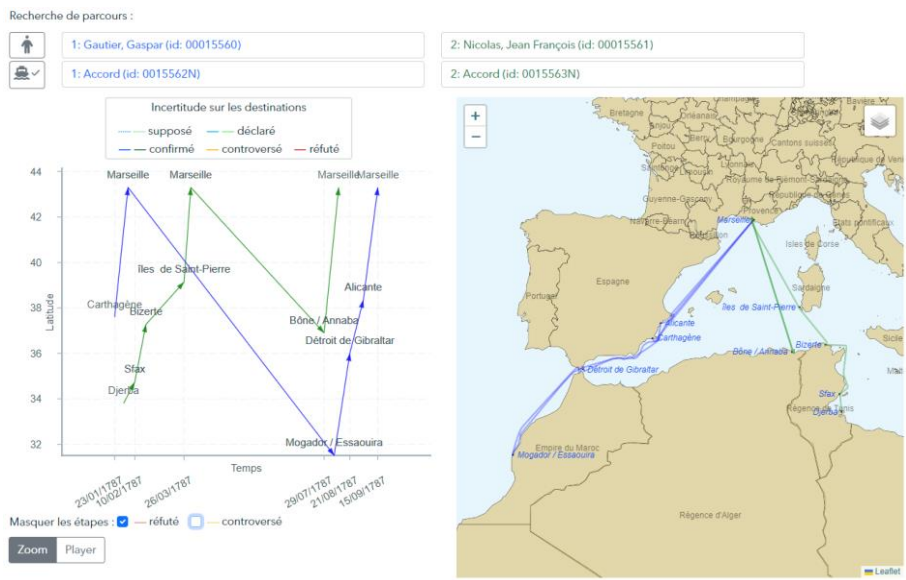

FIGURE 6. *Comparaison d'itinéraires pour deux navires distincts nommés Accord.*

## 5. Conclusion et perspectives

Outre la reconstitution des trajectoires logiques de navires communément identifiés dans différents corpus, l'outil proposé permet d'explorer différents scénarios concernant leur itinéraire. L'approche décrite pour qualifier et représenter des trajectoires maritimes incertaines se révèle utile pour l'histoire maritime. Plus largement, notre approche de contrôle du corpus fondée sur la formalisation de règles expertes interroge notre capacité à expliciter totalement la connaissance experte. L'expertise historienne a été sollicitée plusieurs fois sur des questions incertaines, et cela soulève une question de confiance entre les chercheurs et leurs pairs ou leur public. À plusieurs reprises, les résultats de l'algorithme se sont avérés différents de ceux des historiens, et notre décision a été de le montrer, notamment avec les tronçons qualifiés de controversés dans la reconstitution des trajectoires de navires. Ceci ouvre la piste à de nouvelles analyses historiques.

L'algorithme proposé pour la fabrique de géométries linéaires contournant les côtes pour rejoindre deux escales peut intéresser un autre public. Le code est open-source et modulaire, accessible sur la forge d'Huma-Num[6] tout comme les données sur une API en ligne. Cet algorithme devrait être enrichi prochainement afin de visualiser la navigation sur les fleuves, pour aider à comprendre la circulation sur la Seine de Rouen au Havre par exemple.

---

[6] https://gitlab.huma-num.fr/portic/porticapi pour le modèle de données et l'API,
   https://gitlab.huma-num.fr/portic/shiproutes pour le client Web riche.

Par ailleurs, il peut s'adapter à d'autres bases de données. Le futur musée d'histoire maritime des Sables d'Olonne est intéressé par la représentation interactive de la biographie d'un capitaine sablais du XVIII[ème] siècle. C'est un travail en cours qui a permis tout d'abord de produire des cartes de navigation illustrant la biographie d'un capitaine reconstruite dans une thèse (Retureau, 2020), et illustrant aussi un article de vulgarisation paru dans Olona, une revue de société savante locale spécialisée sur l'histoire des Sables d'Olonne (Chambaud, 2023).

**Bibliographie**

Arnaud, A., Davoine, P.-A., (2009). Approche cartographique et géovisualisation pour la représentation de l'incertitude, in: SAGEO. Paris, France, p. 12.

Bahoken, F., (2021). La représentation graphique de narrations de mobilités spatiales, aspects formels, In: *Comment cartographier les récits / Mapping Méthodologies*, Éditions de l'UMR Territoires, p. 22.

Bochaca M., Moal L. (2019). Le Grand Routier de Pierre Garcie dit Ferrande, Ed. PUR, Rennes, France, 495 p.

Butil, G., (2003). Cabotage et caboteurs de la France méditerranéenn (XVIIe-XVIIIe s.), in: *Rives nord-méditerranéennes* 13, p. 75-91. https://doi.org/10.4000/rives.164

Chambaud, A/, Plumejeaud-Perreau C. (2023). La mer au XVIII[ème] siècle : PORTIC et le numérique. *Olona*, n°263, 2023

Chauveau E., Jégou P., Prcovic N. (2017). Weather Routing Optimization: A New Shortest Path Algorithm. *29th IEEE International Conference on Tools with Artificial Intelligence*, ICTAI 2017, Nov 2017, Boston, United States. hal-01792118

Dedieu, J.-P., Marzagalli, S., Pourchasse, P. Scheltjens, W., (2011). Navigocorpus, a database for shipping information. A methodological and technical introduction, in: *International Journal of Maritime History* XXIII:2, p. 241-262.

Dupuy, M., d'Ambrosio C., Liberti L. (2021) Optimal paths on the ocean. ⟨hal-03404586⟩

Dubucs, H., Dureau, F., Giroud, M., Imbert, C., André-Poyaud, I., Bahoken, F., (2011). Les circulants entre métropoles européennes à l'épreuve de leurs mobilités. Une lecture temporelle, spatiale et sociale de la pénibilité, in: *articulo* 7 https://doi.org/10.4000/articulo.1810

Ducruet, C., (2013). Histoire maritime et cartes en ligne, du XVIe au XXIe siècle. in: *Mappemonde* 109.

Jeansoulin, R., Papini, O., Prade, H., Schockaert, S. (Eds.), (2010). *Methods for Handling Imperfect Spatial Information*, *Studies in Fuzziness and Soft Computing*. Springer Berlin Heidelberg, Berlin, Heidelberg. https://doi.org/10.1007/978-3-642-14755-5

MacEachren, A.M., Robinson, A., Hopper, S., Gardner, S., Murray, R., Gahegan, M., Hetzler, E., (2005). Visualizing Geospatial Information Uncertainty: What We Know and What We Need to Know. in: Cartography and Geographic Information Science 32, p. 139–160. https://doi.org/10.1559/1523040054738936

Marzagalli, S., Pfister, C., (2014). Les pratiques administratives des amirautés du XVIIIe siècle: entre spécificité locale et uniformisation. L'exemple de la gestion des congés, in: Revue d'histoire maritime 19, p. 259-280.

Plumejeaud-Perreau, C., Mimouni, M., Bouju, A., Pfister-Langanay, C., Sauzeau, T., Marzagalli, S., 2021. Un gazetier des places portuaires françaises du xviiie siècle. revuehn. https://doi.org/10.4000/revuehn.1164

Plumejeaud-Perreau et Marzagalli, 2022. Plumejeaud-Perreau, Christine, Marzagalli, Silvia 2022, Qualifying and representing confirmed, possible, invalidated, controversial, inferred eighteenth-century ships' itineraries. Dealing with uncertainty with Navigocorpus database. *Atelier 2. SI pour les Humanités Numériques, conférence INFORSID*, 31 mai-3 Juin 2022, Dijon, France

Poncet-Montanges, A., 2013, Final report of Mapping European Navy, on line: https://mappingeuropeannavy.wordpress.com/2013/05/15/final-report/

Retureau, H., 2020, *Sociétés littorales, gens de mer et activités maritimes dans un port en mutation : l'exemple des Sables-d'Olonne (1747-1866)*. Thèse de l'Université de Nantes.

Wessel, P., and W. H. F. Smith (1996), A global, self-consistent, hierarchical, high-resolution shoreline database, J. Geophys. Res., 101(B4), 8741–8743.

Millié J. (2021) Pierre-Etienne Millié, un marin oublié de Mornac, *de la Seudre à la Charente,* Ed. Société d'Histoire du Canton de Marennes et de ses environs, n°47